\documentclass[prl,twocolumn,superscriptaddress,showpacs,nobalancelastpage]{revtex4}

\usepackage{graphicx}

\begin{document}

\title{\textit{In situ} reduction of charge noise in GaAs/Al$_{x}$Ga$_{1-x}$As Schottky-gated devices}

\author{Christo Buizert}
\affiliation{Kavli Institute of NanoScience, Delft University of Technology, PO Box 5046, 2600 GA, The Netherlands}
\affiliation{Quantum Spin Information Project, ICORP, Japan Science and Technology Agency, Atsugi-shi, Kanagawa 243-0198, Japan}

\author{Frank H.L. Koppens}
\affiliation{Kavli Institute of NanoScience, Delft University of Technology, PO Box 5046, 2600 GA, The Netherlands}

\author{Michel Pioro-Ladri\`{e}re}
\affiliation{Quantum Spin Information Project, ICORP, Japan Science and Technology Agency, Atsugi-shi, Kanagawa 243-0198, Japan}

\author{Hans-Peter Tranitz}
\affiliation{Institut f\"{u}r Angewandte und Experimentelle Physik, Universit\"{a}t Regensburg, Regensburg, Germany}

\author{Ivo T. Vink}
\affiliation{Kavli Institute of NanoScience, Delft University of Technology, PO Box 5046, 2600 GA, The Netherlands}

\author{Seigo Tarucha}
\affiliation{Quantum Spin Information Project, ICORP, Japan Science and Technology Agency, Atsugi-shi, Kanagawa 243-0198, Japan}
\affiliation{Dept. of Applied Physics, University of Tokyo, 7-3-1 Hongo, Bunkyo-ku, 113-8656, Japan}

\author{Werner Wegscheider}
\affiliation{Institut f\"{u}r Angewandte und Experimentelle Physik, Universit\"{a}t Regensburg, Regensburg, Germany}

\author{Lieven M.K. Vandersypen}
\affiliation{Kavli Institute of NanoScience, Delft University of Technology, PO Box 5046, 2600 GA, The Netherlands}
\email{L.M.K.vandersypen@tudelft.nl}

\date{\today}

\begin{abstract}
We show that an insulated electrostatic gate can be used to strongly suppress ubiquitous background charge noise in Schottky-gated GaAs/AlGaAs devices. Via a 2-D self-consistent simulation of the conduction band profile we show that this observation can be explained by reduced leakage of electrons from the Schottky gates into the semiconductor through the Schottky barrier, consistent with the effect of ``bias cooling''. Upon noise reduction, the noise power spectrum generally changes from Lorentzian to $1/f$ type. By comparing wafers with different Al content, we exclude that DX centers play a dominant role in the charge noise.
\end{abstract}


\maketitle

The GaAs/AlGaAs 2-dimensional electron gas (2DEG) has been of unparalleled importance in the field of mesoscopic physics \cite{BeenHout} and has found wide commercial application in High Electron Mobility Transistors (HEMTs) \cite{HEMT}. Today, its unique electronic properties facilitate a variety of important developments, such as spin based quantum information devices \cite{RonaldRev}, Kondo physics \cite{GGKondo}, electron interferometers \cite{Heiblum} and counting statistics \cite{FujiCounting}. In these, and similar experiments, progress is hindered by uncontrolled charge fluctuations in the solid state environment.


Charge noise has been studied both locally by monitoring conductance fluctuations in a quantum point contact (QPC) or quantum dot \cite{YuanLi, Timp, DekkerQPC, Liefrink, Sakamoto,  SmithDX, fujisawaruis, Cobdenchain, MichelLek, KurdakHall}, and on a macroscopic scale using resistance fluctuations in Hall bar structures \cite{KurdakHall, LiGated, Renhallbar}. 
Several charge switching sites have been proposed, either near the 2DEG \cite{DekkerQPC, Liefrink, Sakamoto} or in the remote impurity layer \cite{Timp, KurdakHall, LiGated} and more specifically the DX centers \cite{SmithDX}. The charge switching process has been attributed to electron hopping between trap and 2DEG \cite{DekkerQPC, Liefrink, Sakamoto}, electrons leaking from the split gates through the Schottky barrier \cite{Cobdenchain, MichelLek} or (thermally activated) switching between different sites or configurations within the impurity layer \cite{Timp, KurdakHall,LiGated,SmithDX}. 
Trapping of 2DEG carriers can be excluded as the dominant mechanism since 2DEG density fluctuations are too small \cite{KurdakHall, LiGated}. Switching in the impurity layer is successful in explaining the complex gating behaviour observed in submicron Hall devices by Li \textit{et al.} \cite{LiGated}, whereas gate leakage can explain the stabilizing effect of ``bias cooling'' on Schottky-gated devices \cite{MichelLek}.

Here we present measurements of conductance fluctuations of a QPC with an additional insulated electrostatic top gate that allows us to tune background charge switching \textit{in situ}. The technique has proven successful in reducing charge noise in nine different devices fabricated in three runs on two separate wafers in both Tokyo and Delft, and we believe it to hold universally for GaAs/Al$_x$Ga$_{1-x}$As split-gate devices \cite{Sachrajda}. Furthermore, we examine the mechanism behind this noise reduction, its effect on the noise spectrum, and the nature of the charge traps involved in the switching noise.

\begin{figure}[!t]
\includegraphics[width=3.4in]{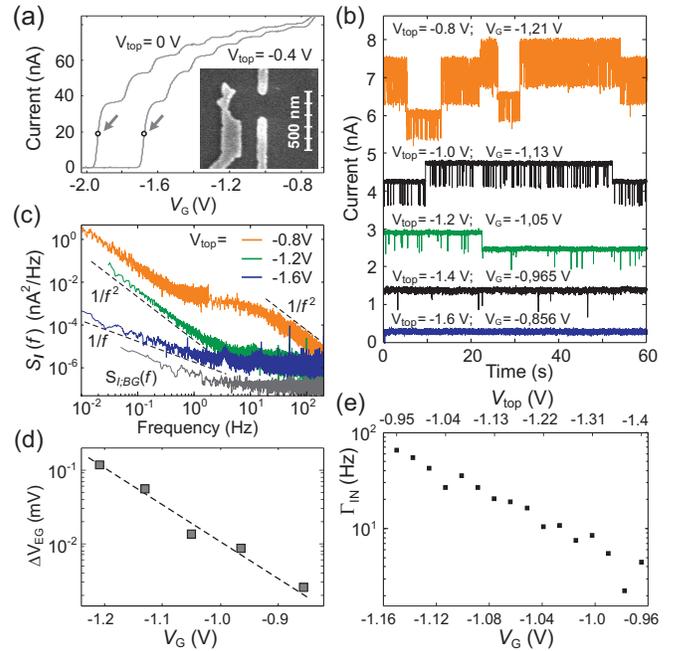}
\caption{Color online (a) QPC pinch off traces (2-terminal, $V_{SD} = 0.8$ mV, $T = 40$ mK). The operating point is marked. Wafer DLF1, see Table \ref{T_Wafers}. Inset: a Scanning Electron Micrograph of a typical device layout before deposition of the insulated top gate. (b) QPC time traces for indicated gate voltages, offset for clarity. (c) Power spectra $S_I(f)$ from FFT of time traces; setup noise background $S_{I;BG}(f)$ recorded at zero $V_{SD}$. (d) Equivalent gate voltage noise $\Delta V_{EG}$. (e) Measured trapping rate $\Gamma_{\rm{in}}$ extracted from time traces as in (b), but for a different QPC.}
\label{F_Vstab_signals}
\end{figure}

A typical device (see Fig. \ref{F_Vstab_signals}a inset) has split Schottky gates (20 nm Ti/Au) deposited on a GaAs/AlGaAs heterostructure with a 2DEG 90 nm below the surface. These gates are covered by a 100 nm thick e-beam defined negative resist calixarene layer, which serves as an electrical insulator \cite{FujitaCalix,calixstab}. Finally we deposit a 400 nm thick Ti/Au top gate with dimensions much larger than the Schottky gates. By applying a voltage $V_G$ to two Schottky gates that approach each other, we deplete the 2DEG underneath and form a QPC. 

Typical QPC pinch-off traces with quantized conductance steps are shown in Fig. \ref{F_Vstab_signals}a, for two values of the voltage applied to the insulated top gate, $V_{\rm{top}}$. We operate the QPC halfway the first plateau ($G_{QPC} \approx e^2/h$) where the slope $dI_{QPC}/dV_G$ is steepest and the signal is most sensitive to changes in the electrostatics. When charge traps close to the QPC are filled or emptied the QPC conductance is modified. The QPC thus provides a local probe of the charge noise.

Fig. \ref{F_Vstab_signals}b illustrates the pronounced effect of $V_{\rm{top}}$ on the charge noise. The topmost trace is very noisy, with several readily-identified two-level fluctuators (from their amplitude, we estimate the charge traps to lie within a few 100 nm from the QPC channel). In the traces below it, $V_{\rm{top}}$ was made more negative in $-0.2$ V increments. Simultaneously the $V_G$ on the Schottky gates was made more positive to maintain $G_{\rm{QPC}} \approx e^2/h$. The fluctuators are eliminated one by one when $V_{\rm{top}}$ ($V_G$) is made more negative (positive).

Figure \ref{F_Vstab_signals}c shows corresponding power spectral densities $S_I(f)$ obtained by Fast Fourier Transform (FFT). The power spectrum $S_I(f)$ of two-level random telegraph noise (RTN) is a Lorentzian which is flat at low frequencies and falls off as $1/f^2$ above the corner frequency $\tau^{-1}_{\rm{eff}} = \tau^{-1}_{u} + \tau^{-1}_{d}$, where $\tau_{u}$ ($\tau_{d}$) is the average time spent in the low (high) current state \cite{Ziel}. For the initial many-level RTN (topmost trace in Fig. \ref{F_Vstab_signals}b), $S_I(f)$ contains Lorentzian contributions with different corner frequencies (topmost trace in Fig. \ref{F_Vstab_signals}c). Once the RTN is eliminated through $V_{\rm{top}}$ the remaining noise has a $1/f$ power spectrum over a wide frequency range, indicative of an ensemble of fluctuators with a homogeneous distribution of timescales $\tau_{\rm{eff}}$ \cite{Ziel}. Also for devices that did not exhibit pronounced RTN at $V_{\rm{top}}=0$ the overall noise level was strongly reduced when a negative $V_{\rm{top}}$ was applied and the QPC was operated at less negative $V_G$.

We quantify the noise level in units of equivalent gate voltage noise $\Delta V_{EG}$, i.e. the voltage noise level applied to the the Schottky gates that would produce the same conductance fluctuations as caused by the charge noise processes. As in Ref. \cite{fujisawaruis} we use the integrated spectral density over a finite frequency range:
\begin{equation}
\Delta V_{EG} = \sqrt{2\int_{0.1}^{100} \left[ S_I(f) - S_{I;BG}(f)\right]df} / \left( \frac{dI_{QPC}}{dV_G}\right)
 \label{E_DV_EG}
\end{equation}
We scale by $dI_{QPC}/dV_G$ to account for device sensitivity. Setup noise $S_{I;BG}(f)$ (see Fig. \ref{F_Vstab_signals}c) is subtracted. We also verified that $S_I(f) \propto V_{SD}^2$, as expected for QPC conductance fluctuations. Fig. \ref{F_Vstab_signals}d shows that $\Delta V_{EG}$ is reduced exponentially with less negative $V_G$.

More insight can be obtained from the $V_{\rm{top}}$ dependence of the RTN timescales. In this case we select a device where a single fluctuator dominates over a relatively large $V_{G}$ range. $V_{\rm{top}}$ is stepped from -0.95 to -1.4 V in increments of -30 mV, while simultaneously $V_G$ is stepped from -1.15 to -0.965 V ensuring $G_{\rm{QPC}} \approx e^2/h$. For each gate voltage setting we record 80 s of the bistable current from which we can derive the trapping and release rates $\Gamma_{\rm{in}} = \tau^{-1}_{u}$ and $\Gamma_{\rm{out}} = \tau^{-1}_{d}$ of the fluctuator. In this $V_{\rm{top}}$ range $\Gamma_{\rm{in}}$ is reduced by over an order of magnitude as shown in Fig. \ref{F_Vstab_signals}e. Whereas this trend in $\Gamma_{\rm{in}}$ is characteristic of all measured devices, changes in $\Gamma_{\rm{out}}$ are generally less pronounced, with both increasing and decreasing trends occurring. Both rates were found to be independent of temperature up to 4.2 K, indicative of tunneling rather than a thermally activated process.

The clear dependence of the RTN on gate voltages, and hence on the conduction band profile below the gates, suggests that its origin is associated with tunnel processes along the growth direction. Specifically, electrons could tunnel from the metal gates through the Schottky barrier to charge traps in the AlGaAs layer ($\Gamma_{\rm{in}}$), and subsequently to the 2DEG ($\Gamma_{\rm{out}}$) \cite{Cobdenchain, MichelLek}.

\begin{figure}[!t]
\includegraphics[width=3.4in]{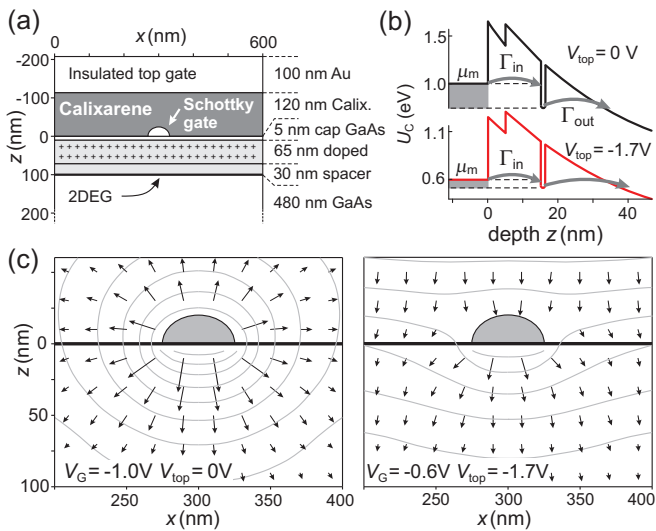}
\caption{(a) Simulated 2-D device structure with Al$_{0.26}$Ga$_{0.74}$As doped- and spacer layer. Si doping $n_{\rm{Si}} = 0.3 \times 10^{18}$ cm$^{-3}$; Calixarene simulated as SiO$_2$ with $\varepsilon_r = 7.1$. (b) Simulated $U_C(z)$ under the Schottky gate at $x=300$ nm. Tunneling into a localized trap with fixed energy below $U_C$ occurs most easily from the quasi-Fermi level in the metal lead ($\mu_m$) where the barrier is lowest (this is generally an inelastic process). (c) Quiver plot of the simulated electric field and equipotential lines near the Schottky gate (grey shaded) for the indicated voltage configurations.}
\label{F_simulationandrates}
\end{figure}

We therefore study in detail how the configuration of gate voltages $\{V_G ; V_{\rm{top}}\}$ affects the conduction band profile $U_C(z)$ and the opacity of the Schottky barrier. To obtain realistic $U_C(z)$ profiles a 1-D calculation would not suffice as the Schottky gate would fully screen changes in $V_{\rm{top}}$. We have performed 2-D self-consistent simulations of our device using the nextnano$^3$ software package \cite{nextnano}. The simulated structure consists of a cross-section of the stacked layers with the Schottky gate embedded in the calixarene insulating layer (Fig. \ref{F_simulationandrates}a).

Here we compare two configurations; the first uses only the Schottky gate to deplete the 2DEG ($V_G = -1.0$ V, $V_{\rm{top}} = 0$ V), whereas the other utilises both gates ($V_G = -0.6$ V, $V_{\rm{top}} = -1.7$ V). These values give identical carrier depletion width (along $x$) at the 2DEG. The corresponding conduction band profiles, $U_C(z)$, directly below the Schottky gate are shown in Fig. \ref{F_simulationandrates}b. We added a possible deep trapping state in the illustration; many such states have been identified in doped AlGaAs quantum well structures \cite{DJ_As}. Clearly the position and energy of the trap influence $\Gamma_{\rm{in}}$ and $\Gamma_{\rm{out}}$. However, for any given trap, the Schottky barrier is higher for more negative $V_{\rm{top}}$ (positive $V_G$). Even though at the surface $U_C(0)$ is always $\approx 0.7$ eV above $\mu_m$ due to surface states, the slope $\partial U_C / \partial z$ is less steep in the lower configuration, and the overall barrier is higher. Furthermore by making $V_{\rm{top}}$ more negative the trap energy is lifted relative to $\mu_m$, reducing the number of allowed initial orbitals (in grey). Eventually leakage is eliminated when the trap energy is lifted above $\mu_m$. In summary, partial depletion using $V_{\rm{top}}$ reduces or even eliminates tunneling from the Schottky gate.

Figure \ref{F_simulationandrates}c shows the 2D electrostatics for both configurations. Note that the radial field in the left configuration also allows tunneling in more sideways directions (possibly also to traps at the surface) and that $\Gamma_{\rm{out}}$ depends on the electric field at the location where the electron is trapped. This can lead to a wide range of behaviors for the influence of $V_{\rm{top}}$ on $\Gamma_{\rm{out}}$, as observed.

\begin{figure}[!t]
\includegraphics[width=2.6in]{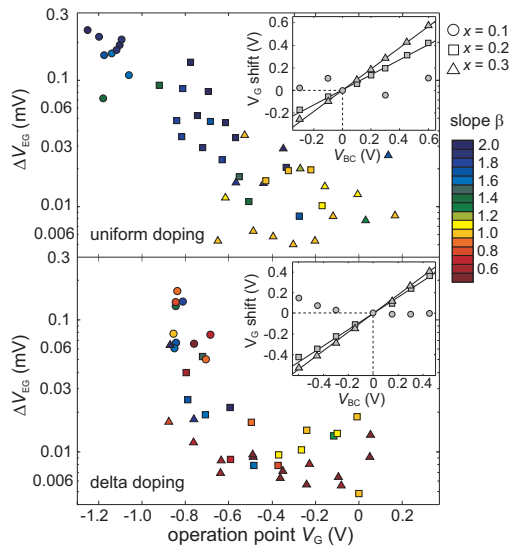}
\caption{Effect of BC on charge noise for uniform doping (REG1-3, upper panel) and delta doping (REG4-6, lower panel). Noise levels calculated using Eq. \ref{E_DV_EG}, the color of the data points codes for the local spectral slope $S_I(f) \propto f^{-\beta }$. The insets show the observed shift in operation voltage $V_G$, at given $V_{BC}$. Each point is averaged over two devices. $T = 4.2$K.}
\label{F_biascooling}
\end{figure}

This interpretation is entirely consistent with the reduction of charge noise due to ``bias cooling'' (BC), see Fig. \ref{F_biascooling} and \cite{MichelLek}. BC is a technique where a device is cooled down with a positive bias $V_{BC}$ applied to the Schottky gates, so that carriers are frozen in at low temperature in deep traps, known as DX centers \cite{Buks}. When $V_{BC}$ is subsequently removed, this (non-equilibrium) trap occupation can be maintained indefinitely. The presence of these additional negative charges lowers $V_G$ required to deplete the 2DEG, as seen in the insets of Fig. \ref{F_biascooling}, and discussed further below.

We here use this BC technique to examine the nature of the traps involved in switching noise. We focus thereby on the question whether DX centers play an important role in charge noise, as is often claimed. For this purpose, we have fabricated split-gate QPCs on six different modulation-doped GaAs/Al$_x$Ga$_{1-x}$As 2DEG heterostructure wafers, where the Al mole fraction $x$ ranged from 0.1 to 0.3 (the dopants were included either uniformly or in a monatomic delta layer, see REG1-6 in Table \ref{T_Wafers}). For $x=0.1$ the trapping energy of DX centers lies on the order of 0.1 eV above the conduction band edge $U_C$ \cite{MooneyDX}, so the DX centers are incapable of trapping electrons (indeed BC does not shift the operation point $V_G$, see Fig.\ref{F_biascooling} inset). For $x = 0.2$ and $x = 0.3$ the DX levels lie around and well below $U_C$ respectively, so they can trap charges during cooldown (additional energy is needed to escape once trapped \cite{MooneyDX}). Thus, for $x=0.2-0.3$, DX centers could potentially act as the intermediate traps responsible for charge noise; for $x=0.1$, they cannot.

Of each wafer a chip with two QPCs was repeatedly cooled down to 4.2 K, each time with different $V_{BC}$. For both QPC's, time traces were recorded with the respective $V_G$ set such that $G_{\rm{QPC}} \approx e^2/h$. Figure \ref{F_biascooling} shows the measured equivalent voltage noise level $\Delta V_{EG}$, as calculated with Eq. \ref{E_DV_EG} for $f =1\ldots 35$ Hz, where each data point corresponds to a single cooldown of a device. The color of the data points codes for the local spectral slope $\beta(f) = -\partial \ln S_I(f) / \partial \ln f$ around $f=10$ Hz.

A clear pattern emerges, despite the large differences in heterostructure composition and the random location of switching sites upon thermal cycling. For very negative $V_G$ we observe high noise levels, often with predominantly Lorentzian spectra ($\beta$ = 2, blue dot), i.e. the signal is dominated by a few active charge traps in the vicinity of the QPC. Via BC we can operate the device at lower $V_G$, leading to systematically reduced noise levels, similar to the case when using $V_{\rm{top}}$. Around $V_G \sim$ -0.4 V the noise level could not be lowered further. Again, the remaining noise is predominantly of the $1/f$-type (yellow dots), and originates from charge noise in the sample. Other noise contributions, such as instrumentation (substracted), $V_G$ noise ($<1 \mu$V peak to peak) and shot noise were all at least an order of magnitude smaller. The delta-doped wafers often showed $\beta<1$ (red dots), indicating a non-uniform distribution of corner frequencies.

Comparing the results for heterostructures REG1-6, we observe that the heterostructures with $x=0.1$ in fact show the highest $\Delta V_{EG}$. Those with $x= 0.3$ exhibit the least charge noise. Furthermore all heterostructures share the common trend of lower $\Delta V_{EG}$ with less negative $V_G$, despite the differences in Al fraction. Based on these observations we exclude the DX center as the dominant trapping site for leaking electrons. Also the suggestion that DX charge state bistability causes the RTN \cite{SmithDX} is refuted. The low band gap energy of Al$_{0.1}$Ga$_{0.9}$As does however make the Schottky barrier more transparent, which might explain why REG4 is much more noisy than REG5,6 for the same $V_G$.

Altogether, we consistently observe that less negative $V_G$ improves charge stability. This can be achieved either by a negative $V_{\rm{top}}$, via BC, or a combination of both. Despite the similarity between the two approaches for noise reduction, it is clear that the resulting electric field profiles are very different, so noise levels and time characteristics may differ even for the same $V_G$. We note that $V_{BC}$ is limited to about +600 mV, which was sometimes not sufficient to stabilize a device, while good stability was achieved with sufficiently negative $V_{\rm{top}}$. The insulated gate approach is thus not only more flexible but proved more powerful as well.

We conclude from these measurements that charge noise in gated GaAs/AlGaAs devices is dominated by electrons tunneling through the Schottky barrier to traps in the AlGaAs layer, where they reside for a while before tunneling out to the 2DEG. After this tunneling mechanism is reduced or eliminated, a baseline charge noise remains, that is presumably of a different origin.
This insight allows us make use of heterostructures that would otherwise suffer from excessive charge noise. It also points to a way of reliably obtaining heterostructure devices with little charge noise, for instance by introducing an additional high bandgap AlAs layer beneath the cap layer, or a thin insulating layer underneath the Schottky gates.

\begin{table}[!t]
\caption{Heterostructure wafer properties}
\label{T_Wafers}
\begin{ruledtabular}
\begin{tabular}{ccccc}
  Name & $x$ & donor depth (nm) & $n$ (cm$^{-2}$) & $\mu$ (cm$^{2}$V$^{-1}$s$^{-1}$) \\
  \hline
  DLF1\footnote[1]{purchased from Sumitomo Electric Industries Ltd., Japan} & 0.27  &  0-70		 &  $4.5 \times 10^{11}$	&  \\
  TOK1\footnotemark[1]  & 0.27  &  5-65		&  $3.0 \times 10^{11}$ & $1.5 \times 10^{6}$ \\
  REG1 & 0.11 &  0-70		&  $1.8 \times 10^{11}$ & $8.5 \times 10^{5}$ \\
  REG2 & 0.2  &  0-70		&  $3.2 \times 10^{11}$ & $2.0 \times 10^{6}$ \\
  REG3 & 0.3  &  0-70		&  $2.8 \times 10^{11}$ & $1.4 \times 10^{6}$ \\
  REG4 & 0.1  &  50	&  $2.1 \times 10^{11}$ & $2.1 \times 10^{6}$ \\
  REG5 & 0.2  &  50	&  $2.3 \times 10^{11}$ & $2.0 \times 10^{6}$ \\
  REG6 & 0.3  &  50	&  $1.3 \times 10^{11}$ & $9.7 \times 10^{5}$ \\
\end{tabular}
\end{ruledtabular}
\end{table}

\begin{acknowledgments}
We thank A. Yacoby, H. Ohno, D.Diggler and the Ensslin group for useful discussions; R. Schouten, B. van der Enden for technical assistance; S. Birner and D. Kupidura for their kind help with nextnano$^3$; R. Ockhorst and P. Rutten for experimental work. L.V. acknowledges financial support by NWO and FOM, W.W by the DFG (SFB 631 and 689) and S.T. by the Grant-in-Aid for Scientific Research S (No.19104007), B (No. 18340081) and Special Coordination Funds for Promoting Science and Technology.
\end{acknowledgments}


\begin{references}
\bibitem{BeenHout}
C.W.J. Beenakker, H. van Houten, Solid State Physics \textbf{44}, 1 (Academic, New York, 1991)

\bibitem{HEMT} R. Szweda, III-Vs Review \textbf{16}, 36 (2003)




\bibitem{RonaldRev}
R. Hanson \textit{et al.}, Rev. Mod. Phys. \textbf{79}, 1217-1265 (2007)

\bibitem{GGKondo}
M. Grobis, et al., Phys. Rev. Lett. \textbf{100}, 246601 (2008)

\bibitem{Heiblum}
D. Chang \textit{et al.}, Nature Physics \textbf{4}, 205 (2008)

\bibitem{FujiCounting}
T. Fujisawa, T. Hayashi, R. Tomita, Y. Hirayama, Science \textbf{312}, 1634 (2006)

\bibitem{YuanLi}
Y.P. Li, D.C. Tsui, J.J. Heremans, J.A. Simmons, Appl. Phys. Lett \textbf{57}, 774 (1990)

\bibitem{Timp}
G. Timp, R.E. Behringer, and J.E. Cunningham, Phys. Rev. B \textbf{42}, R9259 (1990)

\bibitem{DekkerQPC}
C. Dekker \textit{et al.}, Phys. Rev. Lett \textbf{66}, 2148 (1991)


\bibitem{Liefrink}
F. Liefrink, J.I. Dijkhuis, H. van Houten, Semicond. Sci. Tech. \textbf{9}, 2178-2189 (1994)

\bibitem{Sakamoto}
T. Sakamoto, Y.Nakamura, and K. Nakamura, Appl. Phys. Lett. \textbf{67}, 2220 (1995)

\bibitem{Cobdenchain}
D.H. Cobden \textit{et al.}, Phys. Rev. Lett. \textbf{69}, 502 (1992)


\bibitem{SmithDX}
J.C. Smith, C. Berven, M.N. Wybourne, S.M. Goodnick, Surface Sci. \textbf{361/362}, 656 (1996)

\bibitem{fujisawaruis}
S.W. Jung, T. Fujisawa, Y. Hirayama, Y.H. Jeong, Appl. Phys. Lett. \textbf{85}, 768 (2004)


\bibitem{MichelLek}
M. Pioro-Ladri\`{e}re \textit{et al.}, Phys. Rev. B \textbf{72}, 115331 (2005)

\bibitem{KurdakHall}
C. Kurdak, et al., Phys. Rev. B \textbf{56}, 9813 (1997)

\bibitem{LiGated}
Y. Li, et al., Phys. Rev. Lett. \textbf{93}, 246602 (2004)

\bibitem{Renhallbar}
L. Ren, and M.R. Leys, Physica B \textbf{192}, 303-310 (1993)



\bibitem{Sachrajda}
Our results were recently confirmed elsewhere, A.S. Sachrajda, private communication


\bibitem{FujitaCalix}
EPAPS Document no. E-PRLTAO-101-066848. See http://www.aip.org/pubservs/epaps.html.

\bibitem{calixstab} We have observed no improvement in charge stability due to the presence of the calixarene by itself (with $V_{\rm{top}}=0$).


\bibitem{Ziel}
A. van der Ziel, Noise in solid state devices and circuits, Wiley, New York (1986)


\bibitem{nextnano}
S. Birner \textit{et al.}, IEEE Transactions on Electron Devices \textbf{54} (9), 2137 (2007)\
http://www.wsi.tum.de/nextnano3

\bibitem{DJ_As}
D. J. As, P. W. Epperlein and P. M. Mooney, J. Appl. Phys. \textbf{64}, 2408 (1988)


\bibitem{Buks}
E. Buks, M. Heiblum, Y. Levinson, H. Shtrikman, Semicond. Sci. Technol. \textbf{9}, 2031 (1994)

\bibitem{MooneyDX}
P.M. Mooney, J. Appl. Phys. \textbf{67}, R1-R26 (1990)




\end{references}



\end{document}